\documentclass[a4paper,fleqn,usenatbib]{mnras}

\usepackage{newtxtext,newtxmath}
\usepackage[T1]{fontenc}
\usepackage{ae,aecompl}

\usepackage{graphicx}	% Including figure files
\usepackage{amsmath}	% Advanced maths commands
\usepackage{amssymb}	% Extra maths symbols

\title[Heterogeneity of inverted Ca H:K galaxies]{Heterogeneity of Inverted Calcium {\sc ii} H:K Ratio Cluster Galaxies}

\author[K.A.Pimbblet et al.]{
K. A. Pimbblet,$^{1}$\thanks{E-mail: k.pimbblet@hull.ac.uk}
J. P. Crossett,$^{2, 3}$
A. Fraser-McKelvie$^{4}$
\\
$^{1}$E.A.Milne Centre for Astrophysics, University of Hull, Cottingham Road, Kingston-upon-Hull, HU6 7RX, UK\\
$^{2}$School of Physics and Astronomy, University of Birmingham, Edgbaston, Birmingham B15 2TT, UK\\
$^{3}$Instituto de Física y Astronomía, Universidad de Valparaíso, Avda. Gran Bretaña 1111 Valparaíso, Chile \\
$^{4}$School of Physics and Astronomy, University of Nottingham, University Park, Nottingham NG7 2RD, UK\\
}

\date{Accepted XXX. Received YYY; in original form ZZZ}

\pubyear{2018}

\begin{document}
\label{firstpage}
\pagerange{\pageref{firstpage}--\pageref{lastpage}}
\maketitle

\begin{abstract}
The ratio of calcium {\sc ii} H plus H$\epsilon$ to calcium {\sc ii} K inverts as a galaxy stellar population moves from being dominated by older stars to possessing more A and B class stars. This ratio -- the H:K ratio -- can serve as an indicator of stellar populations younger than 200 Myr. In this work, we provide a new method to determine H:K, and apply it to spectra taken of cluster galaxies in Abell~3888. Although H:K is on average systematically lower for the cluster than for a wider field sample, we show that H:K does not have a simple relationship with other indices such as the equivalent widths of H$\delta$ and [O{\sc ii}] beyond having a high value for strong [O{\sc ii}] emission. Moreover, strongly inverted galaxies with H:K $>1.1$ have no preferred location within the cluster and are only slightly lower in their velocity dispersions around the cluster compared to strongly emitting [O{\sc ii}] galaxies. Our results indicate that selecting galaxies on H:K inversion results in a heterogeneous sample formed via a mixture of pathways that likely includes, but may not be limited to, merging spiral galaxies, and quiescent galaxies accreting lower mass, gas rich companions. In concert with other selection criteria, H:K can provide a means to select a more `pure' passive sample or to aid in the selection of highly star-forming galaxies, especially where other spectral line indicators such as H$\alpha$ may not have been observed. 
\end{abstract}


%%%%%%%%%%%%%%%%%%%%%%%%%%%%%%%%%%%%%%%%%%%%%%%%%%

%%%%%%%%%%%%%%%%% BODY OF PAPER %%%%%%%%%%%%%%%%%%

\section{Introduction}

The hierarchical assembly of galaxies in the Cold Dark Matter paradigm unambiguously demonstrates that they build up their stellar mass through gravitational collapse and through repeated merger events with smaller to similar mass objects over their life histories (e.g., White \& Rees 1978; Kauffmann, White \& Guiderdoni 1993; Cole et al.\ 1994; Fakhouri \& Ma 2008; Neistein \& Dekel 2008; Rodriguez-Gomez et al.\ 2015). 
Such mergers
may be `wet' in the sense that a given galaxy accretes gas and dust -- the fuel for star formation -- during these mergers, or they
may be `dry' and simply add pre-existing stellar populations onto themselves (Ciotti, Lanzoni \& Volonteri 2007; Bell et al.\ 2006). Further, they may be `major' mergers (e.g., Bell et al.\ 2006; Kartaltepe et al.\ 2007; Lotz et al.\ 2008; Mundy et al.\ 2017) with a mass ratio of say up to 1:3, or they may be more `minor' (for example, Maller et al.\ 2006; Woods \& Geller 2007). Indeed, up to half of all galaxies have probably experienced a major merger at some point in their history (cf.\ Lin et al.\ 2008) but most will have undergone regular minor mergers since their formation (cf.\ Stott et al.\ 2013; Casteels et al.\ 2014) leading to the conclusion that minor mergers are the dominant type of merger event for all galaxies (Lotz et al.\ 2010).
Although observations show that some galaxies (large ellipticals) have probably formed their stars very early in the history of the Universe and then passively evolved ever since (cf.\ van Dokkum \& Stanford 2003), the natural consequence of such merger processes are multiple stellar populations contained within a given galaxy.  

A major issue in galaxy evolution is how to measure such merger histories observationally (cf.\ Bertone \& Conselice 2009) and build up a picture of how mergers subsequently alter key observational properties of galaxies.
For example, in a relatively wet event, and probably a major merger or a strong tidal interaction, the galaxies involved are likely to trigger a starburst phase that can be seen observationally for $\sim$100's of Myr (e.g.\ Conselice, Bershady \& Gallagher 2000; Di Matteo et al.\ 2008; Overzier et al.\ 2008; Tadhunter et al.\ 2011; Lelli, Verheijen \& Fraternali 2014; but see also Mihos \& Hernquist 1994 for triggering starbursts with minor mergers). 

As starbursts fade away, galaxies enter a transitionary post-starburst period. The cause of this at lower redshifts is   due to rapid quenching thanks to the environment, or a major merger (cf.\ Wild et al.\ 2016). Importantly, this phase provides a crucial link between an energetic star-forming period of a galaxies' life to a quenched and quiescent phase. Typically these two phases are associated with bluer, later-type galaxies for the former, and redder, earlier-type galaxies for the latter. Consequentially, the post-starburst period features a set of observational parameters common to both the types of galaxy that they connect. There are a variety of observational approaches to identifying this post-starburst phase (Wilkinson, Pimbblet \& Stott 2017) and a highly mixed nomenclature for them within the literature depending on how they are selected (cf.\ E+A, k+a, H$\delta$-strong, etc.). Most works seek to eliminate any present-day star-formation through ensuring that spectral features that are associated with on-going star-formation are absent. Most use [O{\sc ii}] and (or) H$\alpha$ for this purpose (e.g., Zabludoff et al.\ 1996; Poggianti et al.\ 1999; Tran et al.\ 2004; Roseboom et al.\ 2006; Poggianti et al.\ 2009; Vergani et al.\ 2010; Rodr{\'\i}guez Del Pino et al.\ 2014). Some authors however do not use H$\alpha$ emission in their selections (Quintero et al.\ 2004; Hogg et al.\ 2006; Goto 2007), whilst others require galaxies to be blue (Dressler \& Gunn 1983; Couch \& Sharples 1987), or use principal component analysis, especially to measure the strength of the break, (Wild et al.\ 2009; 2016).

Almost all use the H$\delta$ absorption line to infer the existence of a frosting of A-class star in the galaxies (although a good number of authors use H$\beta$ and\/or H$\gamma$ as well; for example, see Fisher et al.\ 1998; Tran et al.\ 2003; Blake et al.\ 2004). In combination, these selection criteria provide a method to identify galaxies that were forming stars $\sim$few Gyr ago but have since shut star production down through a quenching mechanism. 

More widely, these observational constraints provide a method through which one can construct a high quality sample of passively evolving cosmic chronometers to provide a determination of the Hubble parameter, $H(z)$, that is independent of cosmology (Jimenez \& Loeb 2002). Such a sample would need to be completely free of younger populations such as the ones discussed above. For galaxy evolution though, these populations are significant due to their ability to connect different phases of the galaxy duty cycle and we can actively select for these younger populations. 

In seeking to assemble samples of passive cosmic chronometers, Moresco et al.\ (2018) provide a highly detailed outline of how to select (against) these younger populations based on a number of widely available observational data that we summarize here. 

Photometrically, a colour-magnitude or colour-colour diagram that incorporates UV light can reveal galaxies that host signs of star-formation from $\sim0.1$ Gyr up to $\sim1$ Gyr ago, and from a very low amount of the stellar mass -- maybe as low as one per cent (see Ferreras \& Silk 2000; Martin et al.\ 2005; Yi et al.\ 2005; Arnouts et al.\ 2007; Kaviraj et al.\ 2007; Schawinski et al.\ 2007; Salim et al.\ 2012; Ilbert et al.\ 2013; McIntosh et al.\ 2014). We took advantage of this method in Crossett et al.\ (2014; 2017) to show that such UV-bright cluster galaxies can be red spirals (cf.\ Masters et al.\ 2010) that are undergoing ram pressure stripping on first cluster infall, and to demonstrate different channels for the quenching of star formation within group environments. 

Spectroscopy yields a number of different diagnostics that can be taken advantage of. Already discussed above are the presence of H$\alpha$ and [OII] lines that are closely associated with recent star formation (say, 10-100 Myr old), and H$\delta$ that is sensitive to A-class stars and therefore, in the absence of the former lines, reveals stellar populations in galaxies that underwent an episode of star formation 0.5-1.0 Gyr ago. 

Arguably one of the most under-used methods in the literature centres around the Ca {\sc ii} K and H absorption lines ($\lambda=3934${\AA} and $\lambda=3969${\AA} respectively). Rose (1985; see also 1984) proposed that the ratio of these lines could be an effective measure of the time since a starburst. The logic behind this is that the H$\epsilon$ line ($\lambda=3970${\AA}) overlaps with the Ca {\sc ii} H line. For galaxies containing stellar classes F, G, and K stars, the ratio of the dip in Ca {\sc ii} H to Ca {\sc ii} K should be almost a constant (see Fig.~6 of Rose 1985). If anything, the Ca {\sc ii} K line will almost always have a stronger dip than the Ca {\sc ii} H line for any galaxy that has a stellar population dominated by older stars (Moresco et al.\ 2018).  However, as a galaxy gains more A and B class stars (or, in general, more massive than approximately F5 class; Rose 1985), the H$\epsilon$ line strengthens and the Ca {\sc ii} lines simultaneously become weaker, resulting in a significantly different ratio -- an \emph{inversion} -- of the H:K ratio. Moresco et al.\ (2018) show that this ratio is sensitive to stellar populations with ages $<200$ Myr -- which is a different timescale to the other indicators discussed above. Although other works have emphasized this ratio (S{\'a}nchez Almeida et al.\ 2012) and demonstrated its importance (Wild et al.\ 2007), it remains a highly under-utilized age indicator, as lower signal-to-noise spectra found in large-scale redshift surveys may not always be able to yield the required measurements.

In this work, we aim to expand our earlier works on post-starburst and frosted populations in galaxy clusters and groups (e.g., Crossett et al.\ 2014; 2017; Pimbblet et al.\ 2006; Pimbblet \& Jensen 2012; see also Mahajan et al.\ 2012) using the H:K ratio diagnostic. Although the H:K ratio has been used on some large spectroscopic surveys, there are very few published works using it within galaxy clusters (e.g., Panuzzo et al.\ 2007 focus on a single cluster galaxy). Hence in this pilot study, we focus on a single galaxy cluster (Abell~3888) and pose some basic questions about this measurement of galaxies: (1) is the H:K ratio for galaxy clusters systematically lower than in the field?; (2) hows does the H:K measurement vary with other spectral indices?; (3) where are the H:K inverted galaxies in the cluster?; and (4) are inverted H:K ratio galaxies an homogeneous population?

Where required, we adopt cosmological parameters corresponding to a flat Universe (Spergel et al.\ 2007): $H_0=$69.3 km s$^{-1}$ Mpc$^{-1}$,  $\Omega_M$=0.238 and $\Omega_\Lambda$=0.762.

\section{Data}
For this, an exploratory work, we choose to work with Abell~3888. This galaxy cluster has been well-studied by ourselves in the past (Pimbblet et al.\ 2002; 2006; Krick, Bernstein \& Pimbblet 2006) and possesses requisite high quality spectra (see Wild et al.\ 2007). As an X-ray luminous cluster ($L_X=10.6 \times 10^{44}$ erg s$^{-1}$ cm$^2$; Reiprich \& B{\"o}hringer 2002), it has also been investigated by a number of other authors who have cross-matched their results with ours to detail the existence of multiple sub-groups within the cluster (Shakouri, Johnston-Hollitt \& Dehghan 2016; Pimbblet et al.\ 2006). Notably, we already have a good handle on the `traditional' post-starburst population within this cluster (Pimbblet et al.\ 2006) defined using [OII] and H$\delta$ and hence are in a position to compare their locations to any H:K inversion galaxies also present. 

The spectra used in this work comes from the Las Campanas and Anglo-Australian Rich Cluster Survey (LARCS; Pimbblet et al.\ 2001; 2002; 2006). The full details of the observations undertaken and their reduction are contained in Pimbblet et al.\ (2006). Here we summarize the most pertinent information about these data. We observe Abell~3888 using the Two-Degree Field (2dF) spectrograph (Lewis et al. 2002) on the 3.9m Anglo-Australian Telescope in three configurations of 2dF, each of $5\times1800$ s, using the 600V grating. This set-up effectively covers the restframe wavelengths of approximately 3600--5400 {\AA} and provides a dispersion of 2.185 {\AA}/pixel with a resolution of $\approx3$ {\AA} in the region of Ca {\sc ii} K and H lines.  The integration time used gives a signal-to-noise ratio (S:N) of $>$15 per pixel in the faintest galaxies in the sample ($R\sim19$). 

Targets for observation come from $R$-band imaging on the 1m Swope Telescope (Pimbblet et al.\ 2001) and are prioritized for observation by 2dF according to both their distance from the cluster core, and their apparent magnitude (see Pimbblet et al.\ 2006), such that bright galaxies (brighter than $M^{\star}+1$) in the core (projected radius $<$30  arcmin) are more likely to be observed than faint ($M^{\star}+1$ to $M^{\star}+3$) core galaxies, which in turn are more likely than bright halo (30 arcmin $<$ projected radius $<$ 60 arcmin) galaxies to be observed. The fourth possible set of faint halo galaxies are not observed. 

Spectra are reduced in a standard manner using the dedicated 2dF reduction pipeline in an identical manner to Colless et al.\ (2001). Redshifts are determined through cross-correlation to a suite of template spectra (Pimbblet et al.\ 2006). 

Finally, throughout this work, we use the completeness function of Pimbblet et al.\ (2006 -- see their Fig.~3) to weight all fractions reported and other relevant quantities. The use of this completeness function is emphasized in the text. 

\subsection{Cluster Membership}
\begin{figure}
\vspace{-1cm}
\hspace{-0.6cm}	\includegraphics[width=3.85in]{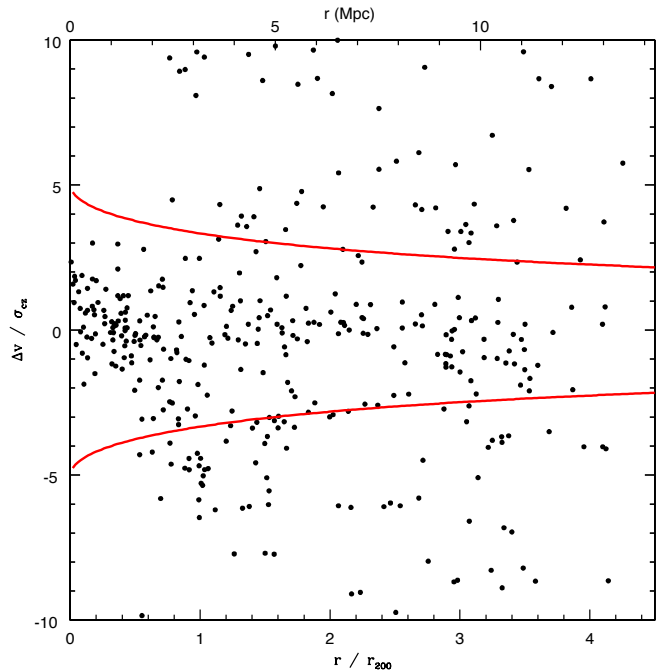}
\vspace{-2.6cm}
    \caption{Cluster membership caustics in the radius-velocity offset phase-space using the mass modelling technique of Carlberg et al.\ (1997). Galaxies between the solid curves ($3\sigma$ caustics) are considered members of the cluster. }
    \label{fig:cye}
\end{figure}

To determine membership of the cluster, we use the prescription of Pimbblet et al.\ (2006) and refer the reader to that publication for in-depth detail. In brief, we apply the mass model of Carlberg et al.\ (1997; see also Balogh et al.\ 1999) that overlays caustics on the phase-space diagram of radius versus velocity offset using a theoretical model that assumes a cluster is a single isothermal sphere. We acknowledge that this assumption will be invalid for almost all clusters, however, the technique yields similar membership outputs compared to other techniques (e.g., Zabludoff, Huchra \& Geller 1990; Diaferio 1999; cf.\ Shakouri et al.\ 2016), and is both convenient, and sufficient, for our needs here at the expense of a small handful of likely infalling interlopers being present (cf.\ Pimbblet 2011; Rines et al.\ 2005). This methodology is illustrated in Fig.~\ref{fig:cye}. Additionally, we note that the velocity centre of the cluster is taken from Pimbblet et al.\ (2006) as 45842$\pm$93 kms$^{-1}$ with $\sigma_z=1328^{+72}_{-62}$ kms$^{-1}$. These values are within $\sim$2$\sigma$ of the same values derived by Shakouri et al.\ (2016). 

% Figure 1 -- Cluster Membership

As a final step, we exclude any galaxy from our sample that has a poor Tonry-Davis value (Tonry \& Davis 1979), or that suffers from  CCD defects in the critical restframe region 3910--3995 {\AA} that are flagged during the data reduction process. In total this leaves 235 galaxies defined as cluster members to work with for this study.

\section{Measuring H:K}
There are multiple ways in which one can measure the Ca {\sc ii} H+ H$\epsilon$ $/$ Ca {\sc ii} K ratio (herein referred to simply as H:K).

Previous works (e.g.\ Rose 1984; see also Rose 1985, Charlot \& Silk 1994; Leonardi \& Rose 1996) elect to simply use the ratio of the counts at the bottom of the troughs. For observations using photographic emulsion this is advantageous since it will not be sensitive to linearization issues or resolution changes. However, measuring the `dip' in this manner does not provide the significance of the line with respect to the continuum. To achieve this, there needs to be a reference level (i.e.\ a measured continuum level) involved, or even an equivalent width measured, yet there remains a question here of where the dip should be measured \emph{with respect to}. There are a number of possibilities, but the choice of this zeropoint is compounded by the presence of the 4000 {\AA} break being in close proximity to the Ca {\sc ii} H line. 

Although there is no formal Lick index (cf.\ Worthy et al.\ 1994), a number of investigators have long proposed measuring both the Ca {\sc ii} K and H lines in summation (Brodie \& Hanes 1986). For example, Serven, Worthey \& Briley (2005) use a blue and red (pseudo-) continuum index of 3806.5-–3833.8 {\AA} and 4020.7–-4052.4 {\AA} to compute the index for these lines (see also, Franchini et al.\ 2010; Brodie \& Hanes 1986).  

In this work, we introduce a new methodology for measuring the dips of each of these lines.  Although a higher wavelength spectral region like 4020-4050 {\AA} may be relatively free of contaminants, any H:K ratio zeropointed to an average in that region would be dependant on the slope of the spectra and the strength of the break, thereby potentially invalidating any inter-galaxy comparison.  We therefore take advantage of our spectral resolution and (arbitrarily) choose the average of the flux in the 3948--3955 {\AA} region to be our zeropoint. This region -- in between the Ca {\sc ii} lines -- is present in all spectra and represents as reasonable a section of spectra as possible to measure the dip from that fulfils our desire to have a zeropoint being both close to the lines in question and largely free from contamination. We explicitly acknowledge that, in extremis, there may plausibly be some contamination from aluminium resonance lines (3944 {\AA} and 3968 {\AA}; Meggers et al.\ 1975) bleeding in to this region, but the amount and the impact on our measures is expected to be very low if present at all (Jaschek \& Jaschek 1995; cf.\ Serven et al.\ 2005).

\begin{figure*}
\vspace{-11cm}
\hspace{-0.6cm}	\includegraphics[width=7.2in]{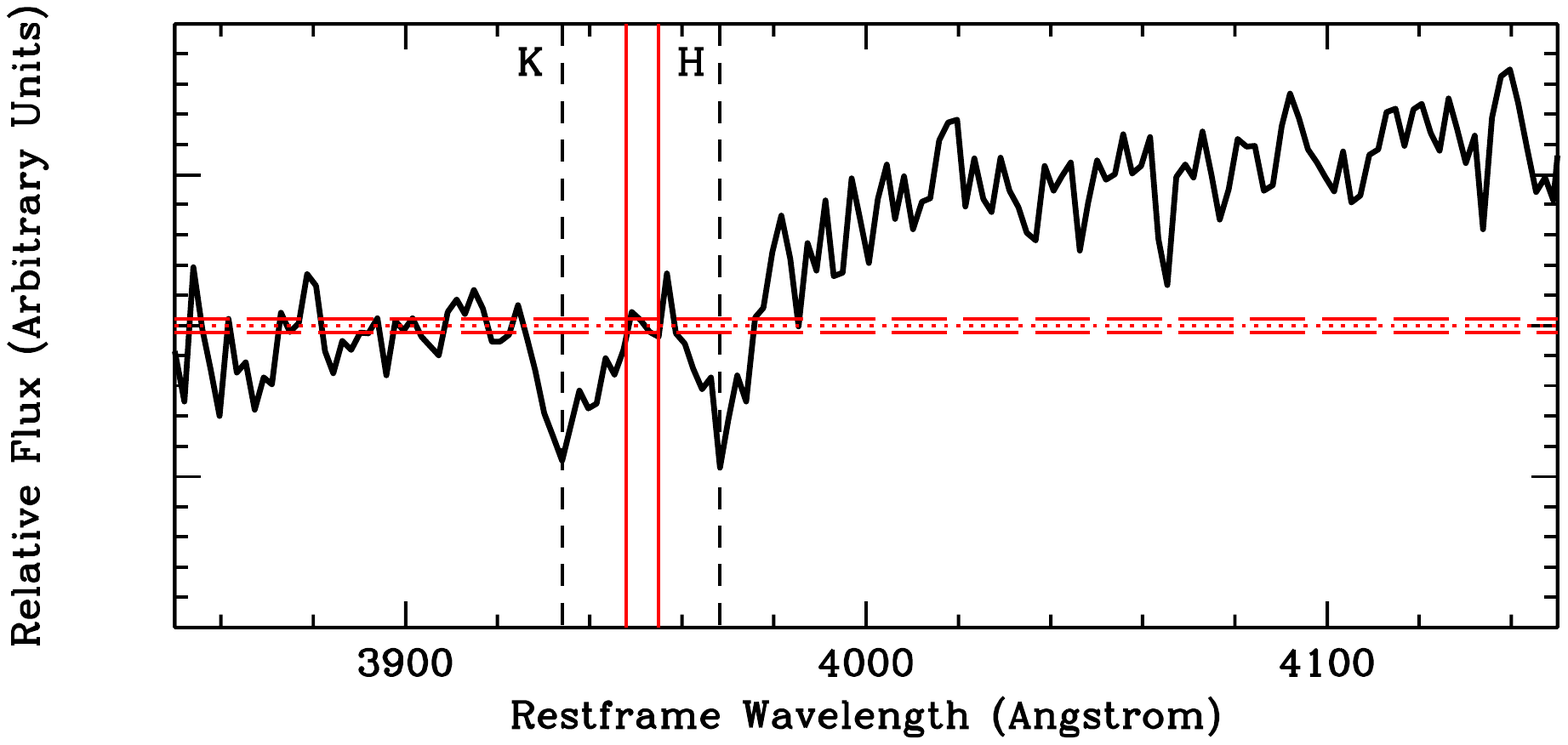}
\vspace{-5cm}
    \caption{An example (unsmoothed) spectrum from our sample to illustrate the H:K computation. The position of the nearest pixel to the Ca {\sc ii} H and K lines are found, and marked here with vertical dashed lines. We measure the `dip' of both of these lines in the spectrum relative to a zeropoint defined as the average flux in the restframe wavelength range 3948--3955 {\AA} (horizontal dotted line). The long dashed horizontal lines show the 1$\sigma$ range on the values used to derive this average flux and we use this as an indicator of the S:N in each of our measured dips. In the example shown, both line dips are more than 20$\sigma$ below the zeropoint and the H:K ratio is 1.05.}
    \label{fig:hkexample}
\end{figure*}

Additionally, we use the standard deviation of the flux of the pixels in this zeropoint region to help give an evaluation of the S:N of the dips measured. Fig.~\ref{fig:hkexample} demonstrates our method on one of our spectra. 

% Figure 2 -- Example of determining H:K

\subsection{Bad Pixels}
To ensure that the adopted method is working as intended, we took the decision to manually inspect all of the spectra. This visual inspection step highlighted a small number of cases where despite the coaddition of multiple observations, dead pixel columns in the data will adversely affect the H:K measurements made due to the pixels in question coinciding with one of the Ca {\sc ii} lines, or being present in the region that the dip measurement is zeropointed to. Whilst the redshift of these spectra are confirmed as correct due to the presence of other lines that are unaffected by such issues, these pixels are flagged as having zero flux in the dataset, which in turn yields an incorrect value for H:K. 

There are two broad options to deal with this issue. One can either discard these spectra as being untrustworthy, or attempt to interpolate over or otherwise remove the individual bad pixels. To preserve the maximum number of spectra under consideration here, we elect to interpolate over the bad pixels whenever possible. This is done by simply replacing the bad pixel values with the average value of the pixels either side of them. In total, 19 spectra have had pixel values interpolated over. Of these, only 2 had the central line in H or K interpolated over, with the rest of the bad pixels featuring in the zeropoint region. 

In undertaking this operation, three spectra showed dead pixels across not only one of the key H or K lines, but in multiple adjacent pixels to them as well. For these spectra, any determination of H:K is therefore not going to be remotely accurate via interpolation and we choose to remove these spectra from the sample.

\subsection{Corrections}
The methodology outlined above introduces a new way to compute H:K. However, we can further refine this. Firstly, regarding the relative spectrophotometricity, we note that there could exist flux tilting at the location of the calcium lines. Such a tilt has the potential to alter our proposed H:K measurement in either direction given that there is nothing in our scheme (or the original Rose scheme) that aims to remove such a tilt. One way to combat this would be to implement flanking pseudo-continua measurements similar to equivalent width Lick indices. However, as noted above, although a blueward continua could readily be defined, a redward one -- wherever placed -- runs the risk of being compormised due to the close proximity of the continuum break. We regard this issue as the primary drawback of our method, and of its antecedents, for which there exists no optimal solution. Here, we note this issue as a significant caveat to our approach.

We turn now to the emission properties of our galaxies. For a minority of our galaxies where we are able to measure a significant H$\beta$ ($\lambda=4861${\AA}) emission, we could compute an emission correction to the H$\epsilon$ line. 
Osterbrock \& Ferland (2006) demonstrate that the ratios for the Balmer lines are fairly static, giving H$\epsilon / $H$\beta = 0.158$ under the assumption that the gas is optically thick (case B recombination) and 10$^4$K. This ratio hardly changes for the optically thin case (where it is 0.159), hence we adopt a ratio of 0.158. We account for the difference in continuum levels between these two lines to produce a delta function that models the H$\epsilon$ emission. This delta function is then smoothed by a Gaussian at the appropriate velocity dispersion and we adopt the peak value of this Gaussian to be the H$\epsilon$ emission correction required. In all cases where we have significant H$\beta$ that is measurable, we determine that the resultant correction to H$\epsilon$ is small -- much less than our uncertainty on the H$\epsilon$ dip described above. Therefore this correction to the H$\epsilon$ line has negligible effect to the results described in the rest of this work.

Finally, we explicitly note here that the combination of the resolution of the spectrograph and the velocity dispersion of the galaxies can strongly alter the depths of the Ca {\sc ii} K and H lines being measured. As an illustration of this point, consider if we were to arbitrarily increase the velocity dispersion of a galaxy. The Ca {\sc ii} K and H troughs would become progressively smoothed, become similar to one another, and approach the central pseudocontinuum level, thus becoming more sensitive and prone to error. Potentially, this is where our redefined method to measure H:K has benefits as we will reject those cases where the lines approach this level. However, it remains the case that there are a range of velocity dispersions of the galaxies in our sample. To get around this and place our measurements on an equal footing, we elect to smooth all of the spectra in our sample to the largest velocity dispersion in our sample.

\subsection{Comparison}
The method we have implemented here does not correspond one to one with the original method of measuring the dip of the Ca {\sc ii} lines presented in earlier works (cf.\ Rose 1984, 1985). In particular, we highlight that this method will produce values that are inverted compared to these works since the former dip method measures the lines to the axis whereas we measure from a mid-continuum level between the lines to their trough. 

\begin{figure}
\vspace{-1cm}
\hspace{-0.6cm}	\includegraphics[width=3.85in]{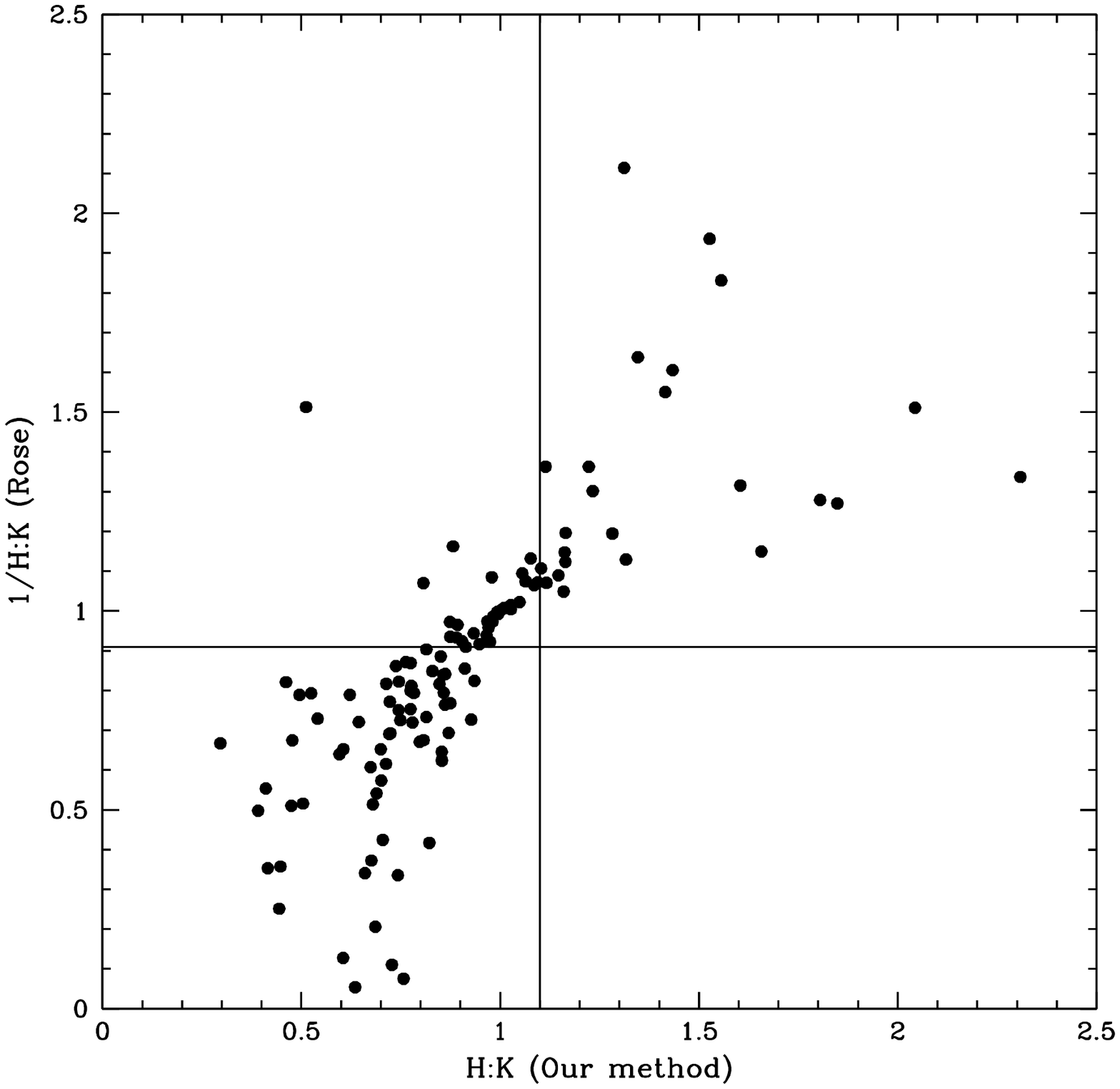}
\vspace{-2.6cm}
    \caption{Comparison of the values obtained for our method versus the former dip method of Rose (1984). The lines correspond to significant `inversion' for each method. }
    \label{fig:HKcompare}
\end{figure}

In Fig.~\ref{fig:HKcompare}, we present a direct comparison between the values for H:K that we obtain versus implementing the earlier method. The vertical line on this diagram shows H:K$=1.1$ -- a value that we herein adopt as constituting an `inverted' galaxy. The horizontal line shows H:K$=1/1.1$ -- i.e. the definition of Moresco et al.\ (2018). 

There are two notable features about this comparison. Firstly, whilst containing scatter the distribution is roughly of a linear form. Secondly, our definition of an inverted galaxy (H:K$>1.1$) is more conservative than Moresco et al.\ (2018).

% Figure 4 -- H:K method comparison 

\section{H:K Values}
From herein, we regard a secure determination of the H:K ratio as only those galaxies whose measured dip in \emph{both} H and K lines are more than three times the standard deviation of the flux in the zeropoint region and an inverted H:K value to be H:K$>1.1$. As demonstrated in Fig.~\ref{fig:HKcompare}, this is a truly conservative defintion of inversion. Excluding the galaxies affected by bad pixel columns, this gives a final sample of 132 cluster members with a secure H:K ratio. This is less than the number that would have been produced via the former dip method where every spectra produces and contributes a H:K value, but we contend these measurements are more secure. 

\begin{figure*}
\vspace{-2cm}
\hspace{-0.6cm}	\includegraphics[width=7.2in]{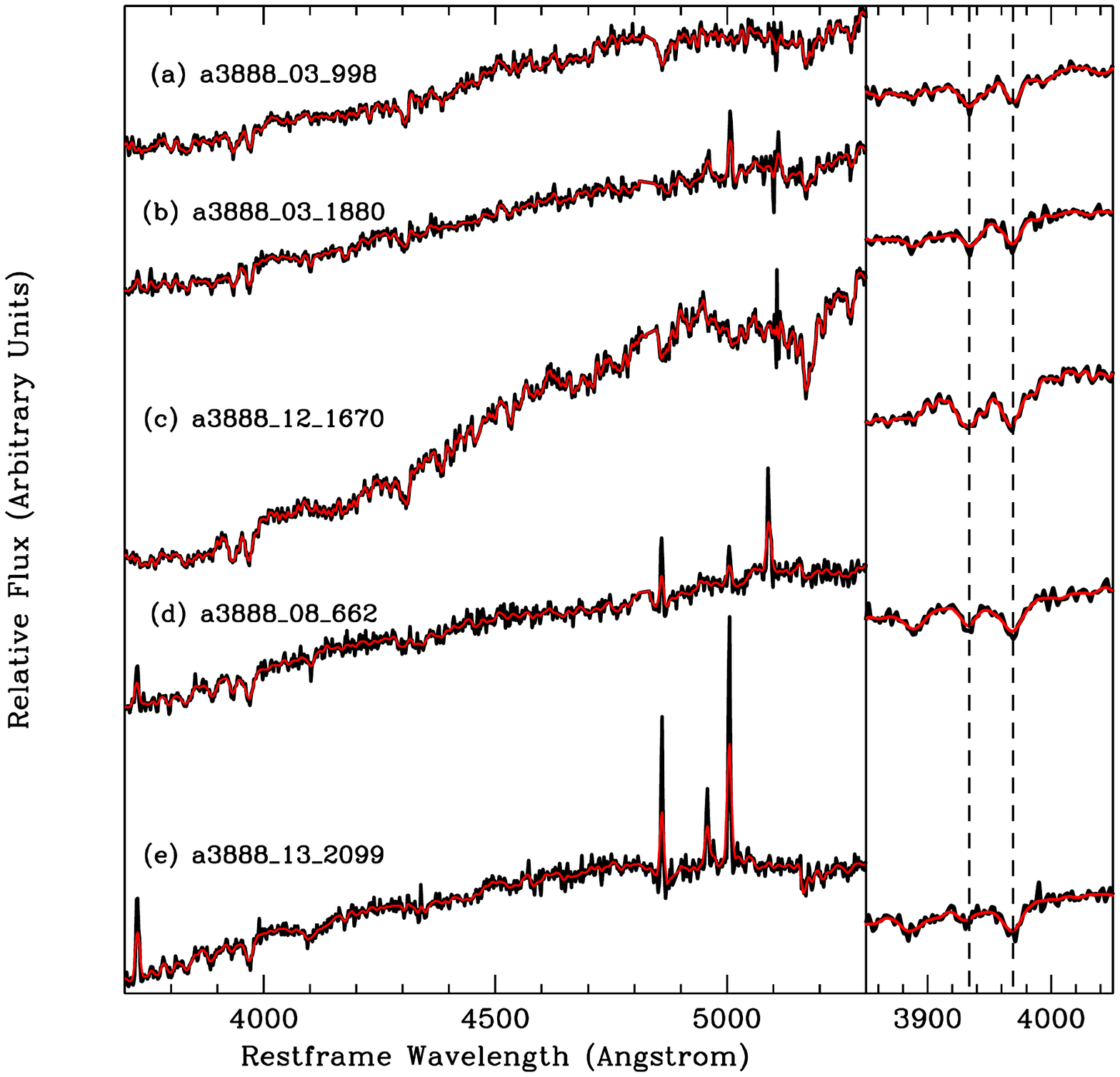}
\vspace{-3cm}
    \caption{Example spectra of five cluster members. The names (a to e) are identical to the catalogue of Pimbblet et al.\ (2006). Only skylines or bad pixels close to H$\beta$ and [O{\sc iii}] have been removed redward of 4200{\AA}.
On the left hand side, we show the wavelength range from approximately [O{\sc ii}] through to Mg. The black line shows the original (reduced, deredshifted) spectra. The red line that is largely interior to the black line shows each spectra blurred to the velocity dispersion of the largest value in the sample, as described above. Several of these spectra show emission lines present in [O{\sc ii}] , [O{\sc iii}], and H$\beta$. On the right hand side is a zoom-in of the Ca {\sc ii} H and K region for each of the five spectra, with vertical dashed lines denoting the position of K and H. 
From case (a) to (e), the H:K ratio changes as 0.45, 0.94, 1.22, 1.43, and 1.66 respectively, representing an increasing significance of the H$\epsilon$ line. 
}
    \label{fig:examples}
\end{figure*}

A selection of the spectra in our sample is displayed in Fig.~\ref{fig:examples}. These examples range from examples where H:K is approximately unity to both extreme ends of the distribution of H:K where one of the lines has a much more pronounced dip than the other. From this, it can immediately be seen that strong emission lines such as [O{\sc iii}] feature in conjunction with the higher values of H:K.

% Figure 3 -- Spectra

\begin{figure}
\vspace{-1cm}
\hspace{-0.6cm}	\includegraphics[width=3.85in]{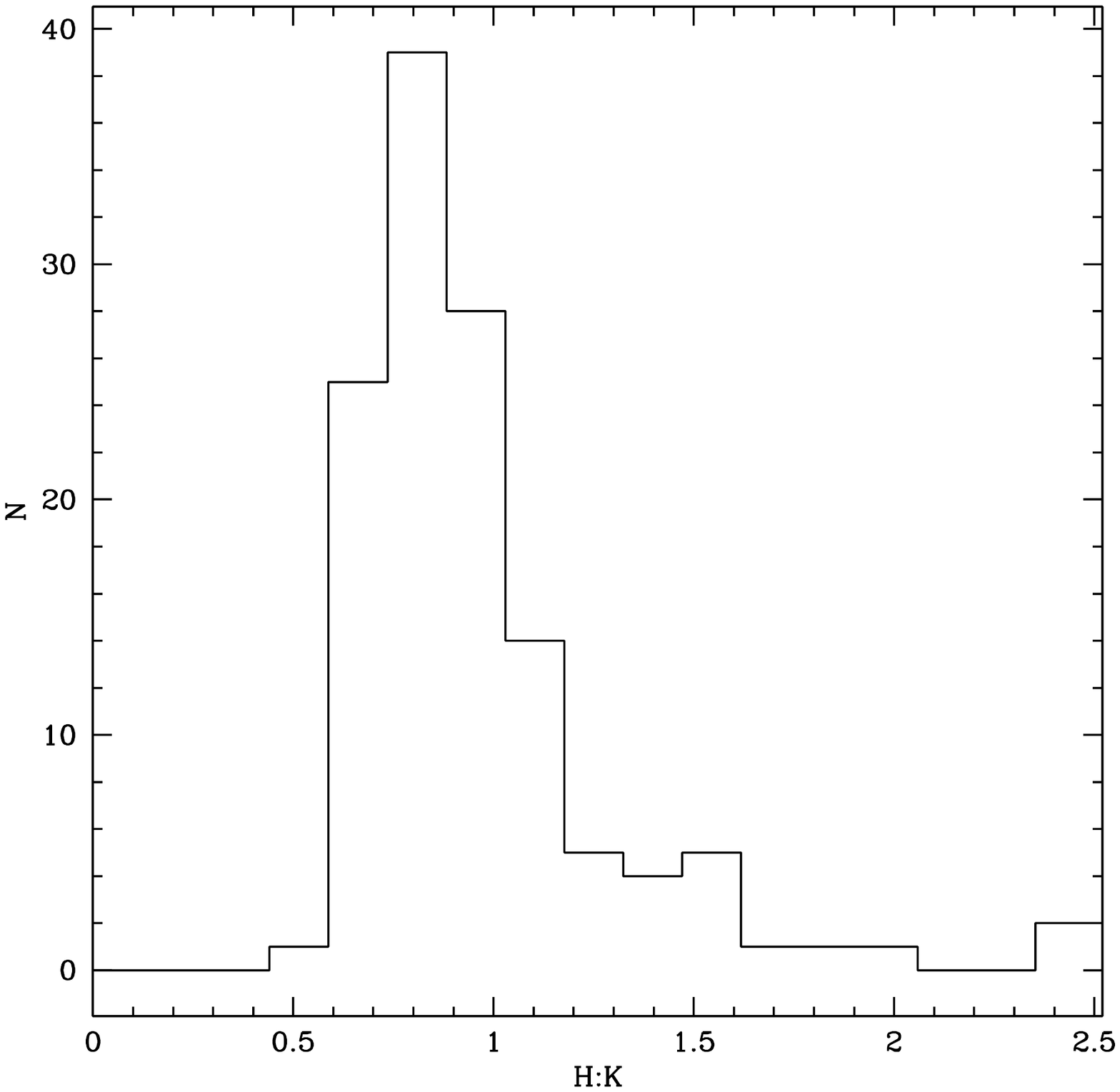}
\vspace{-2.6cm}
    \caption{Histogram of the H:K ratio for all galaxies where the dips in both H and K lines are securely measured. The average value is $<$H:K$>=0.95$.}
    \label{fig:hkvalues}
\end{figure}

The H:K ratio values of our sample is displayed in Fig.~\ref{fig:hkvalues}. The mean value of the H:K ratio in our sample is $<$H:K$>=0.95$, median$=0.87$ (the completeness-weighted mean is also $<$H:K$>=$0.95), with a bootstrapped standard deviation of 0.04. This value is lower than the average value reported by Moresco et al.\ (2018; see also Moresco et al. 2016 and 2012) of $<$H:K$>=$1.17 $\pm$ 0.05. We attribute this to the fact that our sample is composed of galaxies that are members of a cluster.

% Figure 5 -- H:K ratio values 

Moresco et al.\ (2018) use synthetic spectra to explore various levels of young stellar population contamination in galaxy spectra. They report that if a galaxy is to be free of populations $<$200 Myr, then a cut at \emph{their definition of} H:K$=1.1$ could eliminate such populations (again, we emphasize here that our cut is more stringent than this; i.e.\ Fig.~\ref{fig:HKcompare}). In Fig.~\ref{fig:hkvalues}, there is a clear tail of galaxies to larger H:K value bins; representing stronger inversion ratios and therefore younger stellar components.

\subsection{Covariance with [O{\sc ii}] and H$\delta$ Equivalent Widths}
Clearly H:K will probe some of the same regimes of star-formation as other spectroscopic measures. The most obvious line would arguably be H$\alpha$, quantified by its equivalent width value. However, we are unable to determine the covariance of this line with H:K due to the spectral range of our sample and our prior need to ensure that the much lower wavelength [O{\sc ii}] doublet was obtained for our previous publications (Pimbblet et al.\ 2006). 

Indeed, in Pimbblet et al.\ (2006), we used both [O{\sc ii}] and H$\delta$ to classify our cluster samples according to the Poggianti et al.\ (1999) scheme that includes emission line galaxies (e(a), e(b), and e(c) classes), post-starbursts (k+a, a+k classes), and quiescent galaxies (k class). We therefore now ask the question of how both [O{\sc ii}] and H$\delta$ vary with H:K. To do this, we determine the equivalent widths (EWs) of these lines in the same manner as Pimbblet et al.\ (2006). Briefly, we determine the ratio of the flux contained in a passband centred on the index in question to that of the continuum level (i.e.\ the method of Jones \& Worthey 1995). Here, the continuum level is obtained by finding the mean flux in a pair of blue- and redward sidebands. For [O{\sc ii}], the index passband is defined as 3722{\AA}--3735{\AA}, with blue and red sidebands of 3695{\AA}--3715{\AA} and 3740{\AA}--3760{\AA}. Although this index can be sensitive to metalicity and dust, it is a strong indicator of present, on-going star-formation in galaxies (Charlot \& Longhetti 2001). For H$\delta$, there are multiple approaches in the literature as to where to place the passbands (cf.\ Abraham et al.\ 1996 vs.\ Balogh et al.\ 1999) depending if one wishes to capture more flux in the wings of the index in the case of very strong H$\delta$ lines. Although we could in principle take the maximum value of these two approaches (cf.\ Goto et al.\ 2003), we wish to have a consistent definition for our line measurements, notwithstanding its known sensitivity to $\alpha/$Fe variations at supersolar metalicity levels that can lead to incorrect age estimates (cf.\ Thomas, Maraston, \& Korn 2004).  Therefore, here we adopt the H$\delta_A$ definition (Balogh et al.\ 1999) and use an index passband of 4085{\AA}--4124{\AA}, coupled with blue and red sidebands of 4043{\AA}--4081{\AA} and 4130{\AA}--4162{\AA}. 

The flux in the bands is evaluated as :
\begin{equation}
F_{\lambda} = \frac{\sum F_{\lambda , k} / \sigma^2_{\lambda , k} }
{\sum 1 / \sigma^2_{\lambda , k} }
\end{equation}

as per Jones \& Couch (1998) where $F_{\lambda}$ is the summation of the flux for a given index, $\lambda$,
$F_{\lambda , k}$ is flux at $k$ and 
$\sigma_{\lambda , k}$ is the Poisson uncertainty.
The cumulative uncertainty, $\sigma_{\lambda}$, is:

\begin{equation}
\sigma^2_{\lambda} = \frac{1}{\sum 1 / \sigma^2_{\lambda , k} }
\end{equation}

Finally, we determine the EW for the index as:

\begin{equation}
EW = \int_{\lambda_B}^{\lambda_A} (1 - F_{\lambda} / F_{C})
\end{equation}

where $F_{C}$ is the flux of the local continuum.  Following convention, emission lines are presented herein with a negative EW value. 

\begin{figure*}
\vspace{-4cm}
\hspace{-0.6cm}	\includegraphics[width=7.2in]{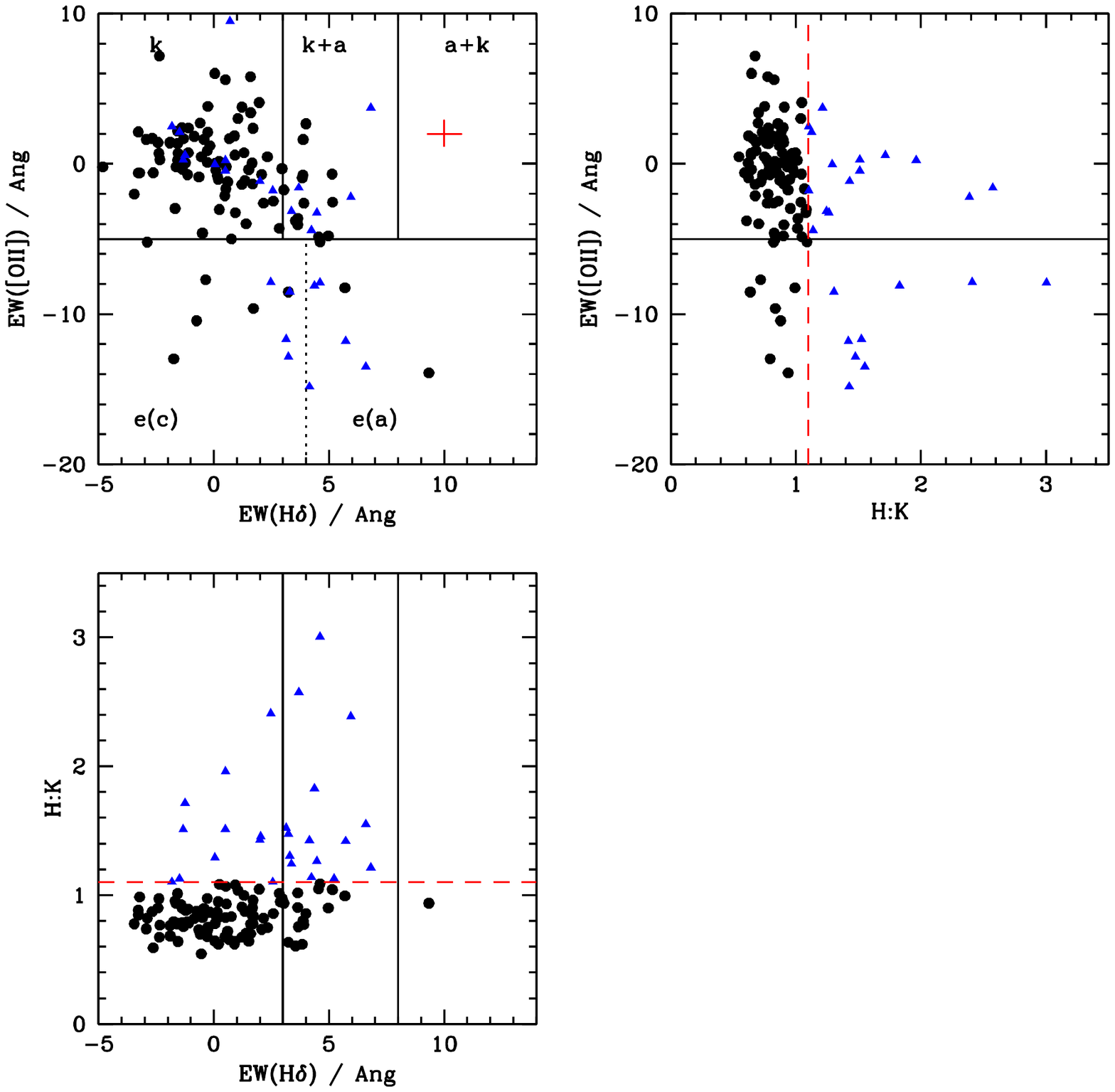}
\vspace{-6cm}
    \caption{{\bf Top-Left:} Cluster members classified according to the Poggianti et al.\ (1999) and Dressler et al.\ (1999) scheme.  The solid horizontal line is located at EW([O{\sc ii}])$=-5${\AA} and delineates the more active, star-forming classes from the more quiescent ones. The vertical lines split the sample on EW(H$\delta$) line strength to give the presence
of A-class stars. Thus the k class in interpreted as quiescent galaxies; k+a and a+k classes (variously labelled E+A galaxies elsewhere in the literature; Zabludoff et al.\ 1996) are post-starburst galaxies depending on how strong the H$\delta$ line is; e(c) class is continual star-forming galaxies and e(a) the same constant star-formation with strong Balmer lines (here distinguished by the dotted vertical line). A further class, e(b), denoting starbursting galaxies is not shown on the figure since we lack any cluster member with the requisite EW([O{\sc ii}]) ($<-40${\AA}). The red cross to the upper right of this panel denotes the typical uncertainty in the EW measurements. Galaxies with H:K$>1.1$ are shown in blue triangles.
{\bf Top-Right:} Relationship between H:K and EW([O{\sc ii}]). The horizontal solid black is in common with the top left panel, whilst the vertical red dashed line denotes H:K$=1.1$ to distinguish strongly inverted H:K ratio galaxies to the right hand side from more normative galaxies on the left following Moresco et al.\ (2018). 
{\bf Bottom-Left:} Relationship between H:K and EW(H$\delta$). The solid vertical lines show the cut between k, k+a, and a+k levels of EW(H$\delta$) with the horizontal red dashed line again denoting H:K$=1.1$. At all levels of EW([O{\sc ii}]) and EW(H$\delta$) there exist galaxies with measurable, and strong H:K inversion (i.e.\ H:K$>1.1$).
}
    \label{fig:hkpog}
\end{figure*}

For EW([O{\sc ii}]), the uncertainties in EW range from a lower value of 4 per cent for quiescent classes range up to 14 per cent in the more emission dominated range $-20${\AA}$<$EW([O{\sc ii}])$<-5${\AA}. For H$\delta$, the uncertainty ranges from 9 per cent in the e(c) range to 24 per cent in the e(a) range (see Fig.~\ref{fig:hkpog}). Examination of the scatter of EW measurements around zero suggests that the modulus detection limit for our lines is approximately 1.5{\AA}. All of these reported values are comparable to Pimbblet et al.\ (2006).

% Figure 6 -- the variation of H:K with Poggianti classifiers (i.e. OII and Hd).

We show in Fig.~\ref{fig:hkpog} the cluster members classified according to the Poggianti et al.\ (1999) and Dressler et al.\ (1999) scheme, in addition to how H:K varies with both EW([O{\sc ii}]) and EW(H$\delta$) in separate panels. The relationship between H:K and EW([O{\sc ii}]) or EW(H$\delta$) is not straight forward. At all levels of both EW([O{\sc ii}]) and EW(H$\delta$) there are galaxies that possess measurable and significant
H:K inversion values -- i.e.\ H:K$>1.1$ (cf.\ Moresco et al.\ 2018). Fundamentally, H:K is probing different regimes (under 200 Myr since star-formation) to either EW([O{\sc ii}]) (which is probing on-going or very recent star-formation) or EW(H$\delta$) (which is probing star-formation in the past 0.5--$\sim$1.0 Gyr). Therefore the lack of direct linear correlation between these values is only mildly surprising. 

A better way to quantify how H:K is changing with these measurements is to determine the completeness-weighted fraction of significant H:K inversion galaxies (i.e.\ H:K$>1.1$) in each class. We present these values in Table~\ref{tab:hkfracs}.
Overall, the completeness-weighted fraction of galaxies with H:K$>1.1$ in the cluster is some 30 per cent. However, when sub-divided in to various samples, the significance of these fractions decreases. In particular, we do not detect any significant trend of the fraction of galaxies having H:K$>1.1$ with EW(H$\delta$) or EW([O{\sc ii}]) -- and by implication from k to k+a to a+k, or from e(c) to e(a). We emphasize that our sample may not be large enough to detect any trend here though.

To probe such changes further and provide indicative ranges of values for each class, we determine the median H:K value for galaxies either side of a EW([O{\sc ii}])$=-5${\AA} cut and a EW(H$\delta$)$=3${\AA} cut. These values are detailed in Table~\ref{tab:hkmedians} along with their inter-quartile range. The values presented in the table supports the inference that all three parameters are probing different star formation timescales and that whilst there is not a simple linear relationship between them.

%
% Table: HK fractions per classification or sample
%
\begin{table}
\begin{center}
\caption{\small{H:K$>1.1$ completeness-weighed fractions in various sub-samples.
}}
\begin{tabular}{lc}
\hline
Sample & H:K$>1.1$ Fraction \\
\hline
All & $0.30\pm0.05$ \\
EW([O{\sc ii}])$>-5${\AA} & $0.29\pm0.09$ \\
EW([O{\sc ii}])$<-5${\AA} & $0.49\pm0.30$\\
EW(H$\delta$)$<3${\AA} & $0.25\pm0.09$ \\
EW(H$\delta$)$>3${\AA} & $0.67\pm0.34$ \\
k      &  $0.27\pm0.09$  \\
k+a  &  $0.58\pm0.40$  \\
e(c) &  $0.36\pm0.35$  \\
e(a) &  $0.71\pm0.65$ \\
\hline
\end{tabular}
  \label{tab:hkfracs}
\end{center}
\end{table}

%
% Table: Median values of H:K
%
\begin{table}
\begin{center}
\caption{\small{Median H:K values and inter-quartile ranges (IQR).
}}
\begin{tabular}{lcc}
\hline
Sample & Median H:K & IQR \\
\hline
EW([O{\sc ii}])$>-5${\AA} & 0.87 & 0.29  \\
EW([O{\sc ii}])$<-5${\AA} & 0.94 & 0.48 \\
EW(H$\delta$)$<3${\AA} & 0.87 & 0.29 \\
EW(H$\delta$)$>3${\AA} & 1.03 & 0.54 \\
\hline
\end{tabular}
  \label{tab:hkmedians}
\end{center}
\end{table}

\section{Cluster Location}

%%% FIgure 7. Physical locations of the sub-types within the cluster.
\begin{figure*}
\vspace{-3cm}
\hspace{-0.6cm}	\includegraphics[width=6.2in]{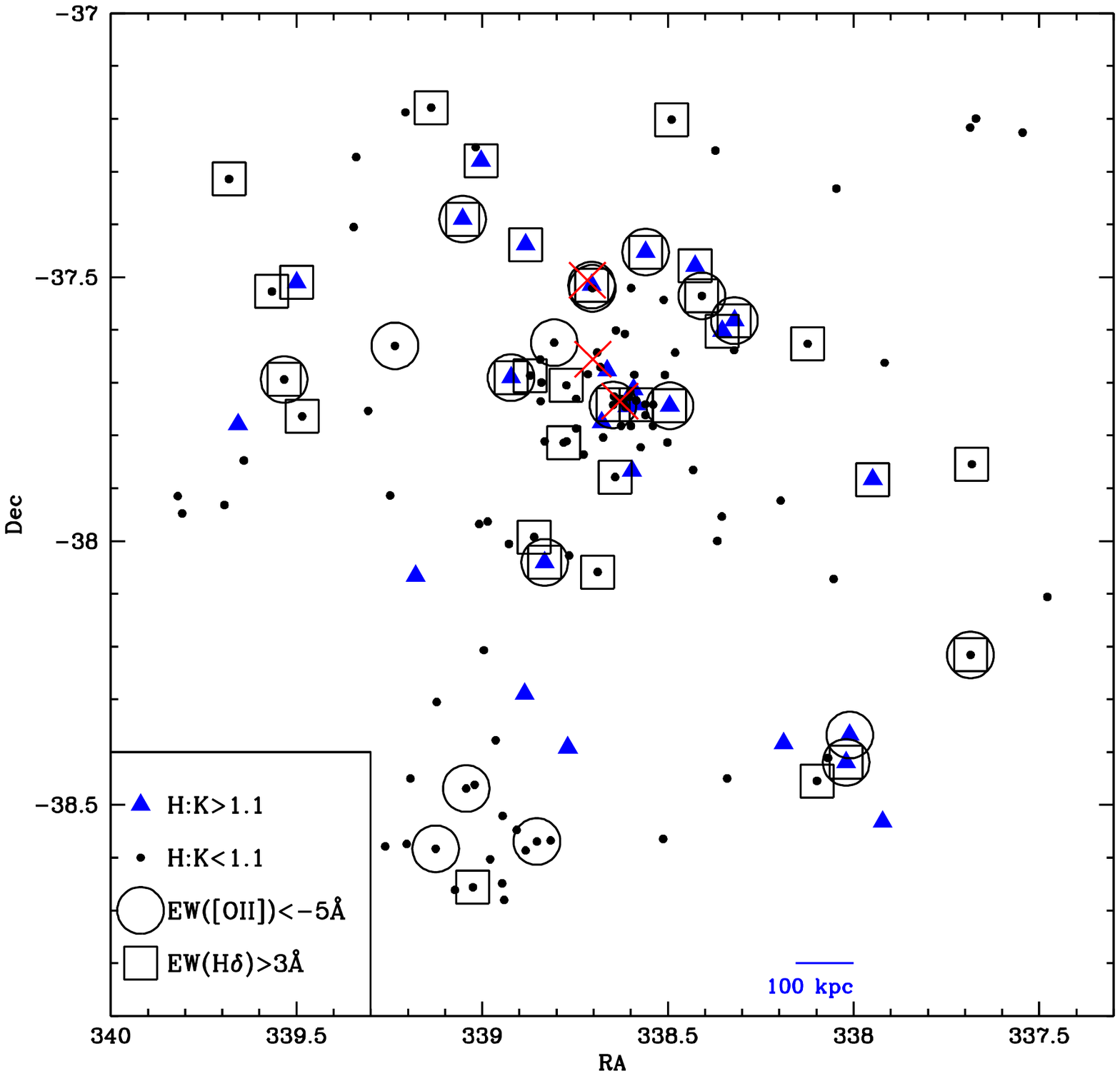}
\vspace{-5cm}
    \caption{Physical locations of galaxies split on H:K (blue triangles above H:K=1 and black dots below). Those with EW([O{\sc ii}])$<-5${\AA} are contained within a large circle, and those with EW(H$\delta$)>3{\AA} are contained in squares. The locations of the known sub-structure within Abell~3888 from Shakouri et al.\ (2016) are denoted by red crosses. There is no preference for any galaxy type occupying the outlying sub-clusters. }
    \label{fig:locations}
\end{figure*}

%%% Figure 8. Phase Space locations of the sub-types.
\begin{figure*}
\vspace{-3cm}
\hspace{-0.6cm}	\includegraphics[width=6.2in]{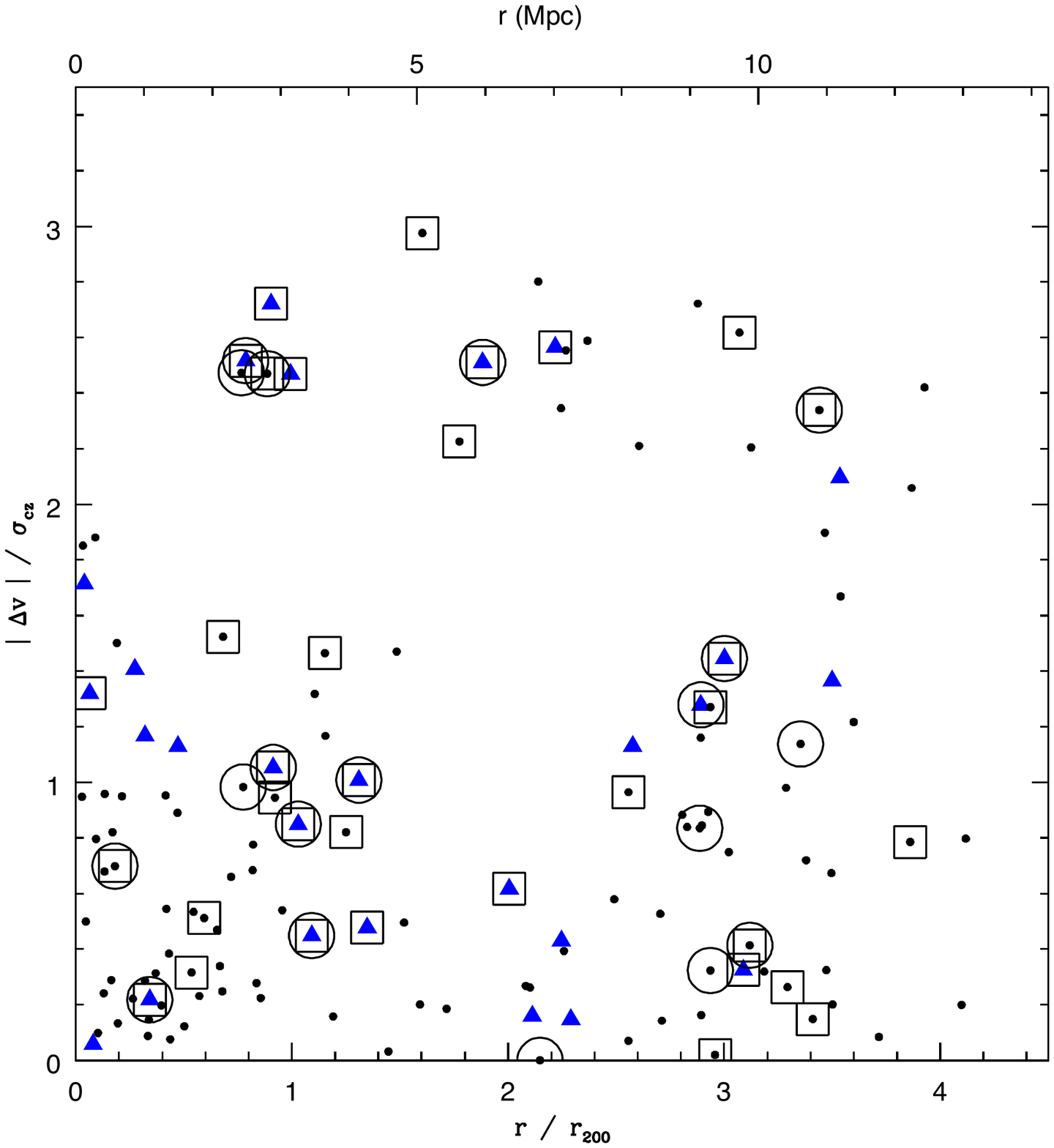}
\vspace{-5cm}
    \caption{Phase space diagram of galaxies split on H:K ratios. The point style is the same as for Fig.~\ref{fig:locations}.}
    \label{fig:cyetyped}
\end{figure*}

We now turn to the question of where the galaxies with strong H:K inversion ratios (i.e.\ H:K$>$1.1) are located inside Abell~3888. In Fig.~\ref{fig:locations} we display a spatial plot of these galaxies, alongside the locations of those possessing significant EW([O{\sc ii}]) and EW(H$\delta$), as well as the locations of the identified sub-clusters taken from Shakouri et al.\ (2016). There does not appear to be any preference for these types of galaxies to occupy the sub-structures indicated. Additionally, the strongly H:K inverted galaxies seem to be spread across the entirety of the cluster. 

To investigate this further, we show in Fig.~\ref{fig:cyetyped} a phase space diagram of the modulus value of the velocity difference of galaxies to the cluster mean normalized by the cluster velocity dispersion, $\Delta V / \sigma_{cz}$, against cluster-centric radius. Once again, the galaxies with H:K$>1.1$ appear to be scattered across the diagram whilst the strongly star-forming galaxies with EW([O{\sc ii}])<-5{\AA} preferring the cluster outskirts.

We quantify this in Table~\ref{tab:veldisps} where we present the median velocity dispersion values of various classes of galaxies normalized against the most passive galaxies in our set (i.e.\ one with H:K$<1.1$, EW([O{\sc ii}]) $>-5${\AA}, and EW(H$\delta$)$<3$). These dispersions are, of course, in line with Pimbblet et al. (2006) for galaxies with high and low values of EW([O{\sc ii}]) and EW(H$\delta$). Remarkably, galaxies that are either side of the adopted H:K$=1.1$ dividing line are very similar in this phase-space, and, in turn, reinforces the visual conclusion from Fig.~\ref{fig:locations} and Fig.~\ref{fig:cyetyped}: they have no preferred location within the cluster. Secondly, they have dispersion values that are only slightly below the high EW([O{\sc ii}]) emission sample. 

%
% Table: Velocity Dispersions, normalized
%
\begin{table}
\begin{center}
\caption{\small{Median velocity dispersion values normalized to H:K$<1.1$, EW([O{\sc ii}])$>-5${\AA}, plus EW(H$\delta$)$<3${\AA} galaxies.
}}
\begin{tabular}{lc}
\hline
Sample & $\sigma(\Delta v / \sigma_cz)$ \\
\hline
H:K$<1.1$, EW([O{\sc ii}])$>-5${\AA}, EW(H$\delta$)$<3$ & 1.00 \\
H:K$<1.1$ & 1.02 \\
H:K$>1.1$ & 1.33 \\
EW([O{\sc ii}])$>-5${\AA} &  1.05 \\
EW([O{\sc ii}])$<-5${\AA} & 1.36  \\
EW(H$\delta$)$<3${\AA} & 1.06 \\
EW(H$\delta$)$>3${\AA} & 1.15 \\
\hline
\end{tabular}
  \label{tab:veldisps}
\end{center}
\end{table}

\section{Discussion}

\subsection{Metallicity}
Metallicity has the potential to affect the Ca {\sc ii} lines on its own. Although it is very unlikely to drive an inversion in older stellar populations (Leonardi \& Rose 1996), younger stellar populations can be inverted for a fixed age. 

Given the shortened wavelength range of our spectra (e.g., they do not contain H$\alpha$), we do not attempt full spectral fitting of the spectra to probe this, nor is the S/N at the fainter magnitudes able to yield reliable stellar metallicities. However, we are able to use other methods to try to obtain a handle on the range of gas phase metallicities present in our sample. In particular, the R23 measurement (Pagel et al.\ 1979; see also e.g.,
Pagel, Edmunds \& Smith 1980;
Dopita \& Evans 1986;
Charlot \& Longhetti 2001; 
Kewley \& Dopita 2002; 
Kobulnicky \& Kewley 2004) is well-suited to our spectra.
In brief, R23 is defined as ([O{\sc ii}] $\lambda$3727 $+$ [O{\sc iii}] $\lambda$$\lambda$4959, 5007)$/$H$\beta$. This measurement is sensitive to abundance, however it has one significant issue associated with its use: it is a double-valued function of abundance (cf.\ Fig.~5 of Kewley \& Dopita 2002). 

Ideally, we would have followed the Kewley \& Dopita (2002) method involving [NII] to break this degeneracy. However, along with H$\alpha$, this diagnostic line is not available to us and therefore we are forced to use  approaches in common with high-z galaxy studies where such lines have been redshifted out of the spectroscopic wavelength observation limits (cf.\ Nagao, Maiolino, \& Marconi 2006). 

We therefore follow the suggestion of Maier, Meisenheimer \& Hippelein (2004; fully confirmed by Nagao et al.\ 2006) that the O32 ratio ([O{\sc iii}] $\lambda$5007 $/$ [O{\sc ii}] $\lambda$3727) can distinguish between the upper- and lower-branches of R23 versus metallicity relationship (explicitly: O32$<2$ indicates that the observed R23 value is an upper-branch value) to break the inherent degeneracy.

We find that we have 8 cluster galaxies with significant and measurable flux in emission in [O{\sc ii}] $\lambda$3727, [O{\sc iii}] $\lambda$5007, and H$\beta$ (where [O{\sc iii}] $\lambda$4959 is not measured, we use an assumed value of [O{\sc iii}] $\lambda$$\lambda$4959, 5007 $= 1.327$ [O{\sc iii}] $\lambda$5007; as per Nagao et al.\ 2006). All of these galaxies belong to the upper-branch of the R23 sequence according to their O32 values.  

For these 8 cluster galaxies with strong emission (i.e.\ young stellar populations), we determine a mean R23$=2.1$ with an inter-quartile range of 0.3--4.6.
From the relation between R23 and metallicity presented by Nagao et al.\ (2006; see their Tables 5 and 6), this results in a mean metallicity of $12+$log$_{10}$(O$/$H)$=9.1$ with a corresponding inter-quartile range of 8.8--9.6 for these galaxies. This metallicity is approximately solar metallicity and higher (i.e.\ $12+$log$_10$(O$/$H)$= 8.69$; Nagao et al.\ 2006). Therefore whilst metallicity may contribute to the H:K inversion for a minority of star-forming or recently star-forming galaxies (cf.\ Figures 3 and 4 of Moresco et al.\ 2018) it is not the sole driver, nor (likely) the primary driver given the above analysis using R23. We re-emphasize here that the modal galaxy in our sample is passive and will therefore be unaffected by such metallicity considerations.

\subsection{Heterogeneity}
\begin{figure}
\vspace{-1cm}
\hspace{-0.6cm}	\includegraphics[width=3.85in]{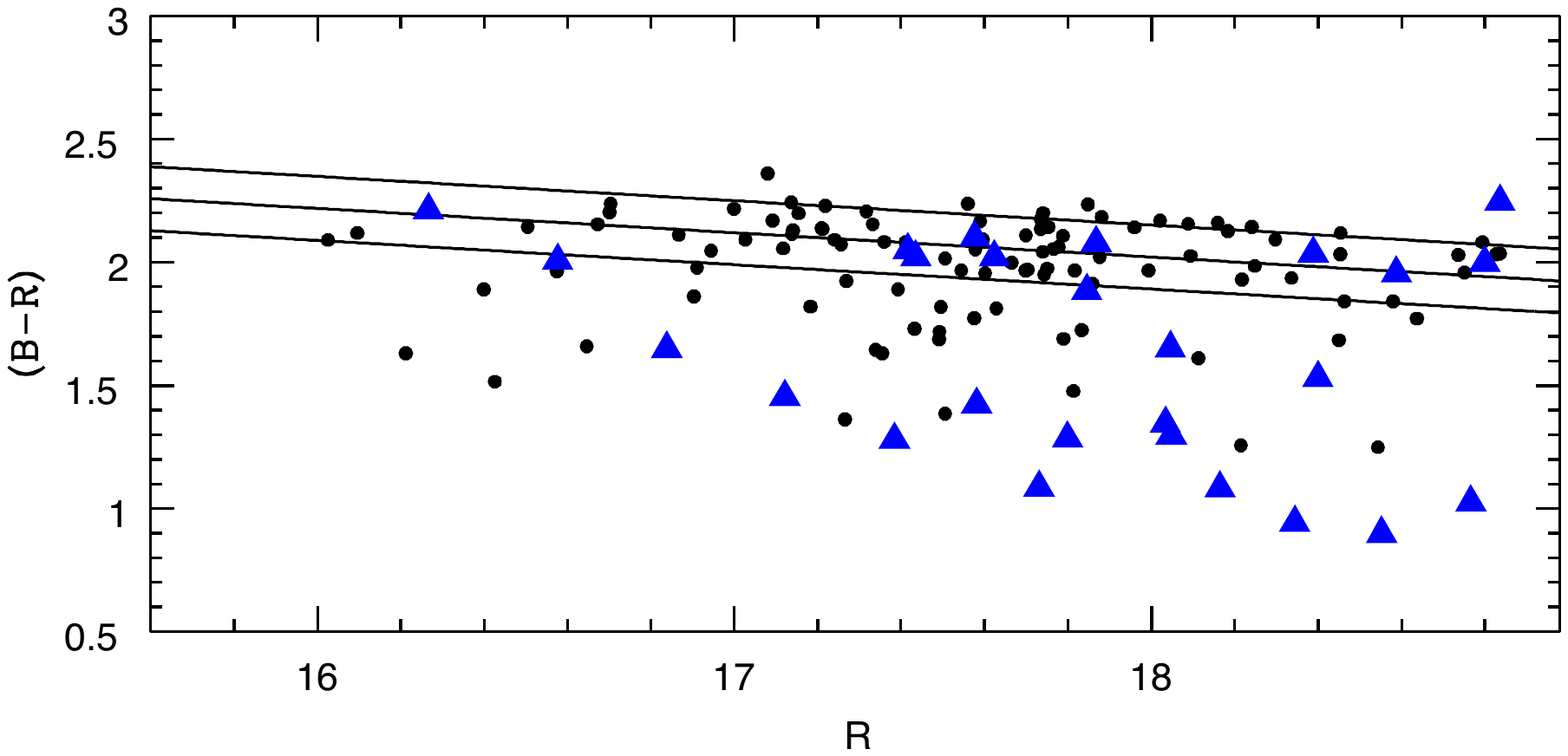}
\vspace{-7.6cm}
    \caption{Colour-magnitude diagram for cluster members with measured H:K values. Photometry is taken from Pimbblet et al.\ (2001) and the sloped lines denote the centre and locus of the colour-magnitude relation taken from Pimbblet et al.\ (2002).  Five galaxies with poor $B$-band aperture photometry are excluded from this diagram. Inverted galaxies with H:K$>1.1$ (blue triangles) are found in all regions of the diagram.}
    \label{fig:colmag}
\end{figure}

One possibility that may explain the, in part, the values of H:K in Table~\ref{tab:veldisps} and the lack of preferred position within the cluster (at least in comparison to other sub-samples) is that selecting galaxies based on their H:K ratios results in an heterogeneous sample. Such heterogeneity is present in other samples as the selection of E+A galaxies demonstrates: it is not sufficient to simply select H$\delta$ strong galaxies to infer post-starburst status -- a lack of [O{\sc ii}] or H$\alpha$ emission must also be present to confirm such a status, as otherwise one may well be selecting e(a) class galaxies alongside k+a's (Dressler et al.\ 1999; Poggianti et al.\ 1999). This is reinforced by several lines of reasoning. In Fig.~\ref{fig:colmag} we display a colour-magnitude diagram for the galaxies with measured H:K values and find that H:K$>1.1$ galaxies are present in all parts of the diagram. To quantify this, we determine that the completeness-weighted average colour for galaxies with H:K$>1.1$ is $(B-R)=1.77$, compared to $(B-R)=1.88$ for galaxies with H:K$<1.1$. The difference between these two colours is the less than three times the uncertainty on the photometry from Pimbblet et al.\ (2001), and substantially smaller at faint magnitudes, not withstanding the gradient in the colour-magnitude relation. As for the morphology of the galaxies, there is little difference either side of the H:K$=1.1$ cut either, with a completeness-weighed concentration index (Abraham et al.\ 1994; 1996; see also Pimbblet et al.\ 2001) of 0.37 for H:K$<1.1$ and 0.35 for H:K$>1.1$.

% Figure 9 -- Colour-Magnitude Diagram.

What then does the measurement of H:K add to an analysis of spectral types and galaxy evolution? Firstly, it provides a critical extra method to ensure the selection of passive galaxies alongside other, more widely used techniques. These are key to various cosmic chronometry experiments (Moresco et al.\ 2018), and the lack of a significant inversion has implications for any study wishing to determine the status and history of truly passive, quiescent galaxies (cf.\ Fraser-McKelvie et al.\ 2016; 2018). As an illustrative example, in Table~\ref{tab:amelia}, we present the H:K values for the six candidate passive spiral galaxies from Fraser-McKelvie et al.\ (2016 - see in particular their Fig.~3). All of these values are below H:K$=1.1$ and hence would not be considered to be significantly inverted by the definition used in this work. Interestingly, the majority of the H:K values are centred around unity rather than being far below. This analysis reinforces the conclusion of Fraser-McKelvie et al. (2016) that morphological transformation is not required to terminate star-formation within a spiral galaxy.

%
% Table: H:K values for the passive sp sample from Fraser-McKelvie et al. (2016).
%
\begin{table}
\begin{center}
\caption{\small{H:K values from the integrated spectra of the six passive spiral galaxy candidates presented in Fraser-McKelvie et al.\ (2016). 
}}
\begin{tabular}{lc}
\hline
Galaxy & H:K \\
\hline
NGC~4794 & 0.97 \\
NGC~538 & 0.99 \\
NGC~4305 & 0.97 \\
IC~375 & 0.87 \\
CGCG~394-006 & 1.00 \\
MGC-02-12-043 & 1.02 \\
\hline
\end{tabular}
  \label{tab:amelia}
\end{center}
\end{table}

At the other extreme, the presence of inverted H:K ratios can be combined with other spectral measures (e.g., EW([O{\sc ii}])) to facilitate selection of highly star-forming galaxies -- especially in the absence of H$\alpha$ where one is reliant on features and indices at lower wavelengths (e.g., [O{\sc iii}], H$\beta$, [O{\sc ii}], etc.).

In galaxies that may be otherwise determined to be passive, the presence of an inverted H:K ratio serves as a signpost to its recent star-forming past that might otherwise go undetected through the use of classifications derived from more common analyses such as [O{\sc ii}] in concert with H$\delta$ (and/or further Balmer lines; cf.\ Blake et al.\ 2004; Wilkinson et al.\ 2017) and may offer an additional method or cut to apply to a sample to identify a transitioning post-starburst galaxy (cf.\ Pawlik et al.\ 2018; Wild et al.\ 2009). 

Naturally, given such heterogeneity, the origins of inverted H:K ratio galaxies must be plural. Consider a spiral and star-forming galaxy that undergoes a violent interaction with a $~\sim$similar mass galaxy. Not only will morphological deformation occur, but a starburst is very likely to result as well. This may transition to a passive post-starburst k+a phase via an emission line phase (one that combines a strong H$\delta$ line alongside detectable emission lines indicative of on-going star-formation such as H$\alpha$ and H$\beta$); analogous to the e(a) phase. Along the earlier part of this sequence, an inversion of the H:K ratio is very likely to be present and persist as more A and B class stars are present.

The second clear pathway would be for a galaxy that is already passive to suddenly undergo a somewhat weak starburst -- most likely due to the impact and merger of a very gas-rich galaxy of lower comparable mass. The transition for this galaxy would then be to go from passive, to being temporarily starforming via a weak starburst through to a post-starburst phase before resuming its past quiescent nature. 

Our analysis here is limited by the absence of H$\alpha$ from our spectra. It may yet be the case that there is a correlation with an active galactic nuclei (AGN) phase. Indeed, given that any trigger capable of causing a starburst can also be likely to cause an AGN phase as well, it is plausible that an inverted H:K ratio may -- but certainly not always -- herald a later AGN phase (cf.\ Cid Fernandes et al.\ 2004; Yan et al.\ 2006; Sell et al.\ 2014; Pawlik et al.\ 2018; see also Pimbblet et al.\ 2013 who point out the issue that the site of any interaction that triggers AGN in clusters is likely long since obfuscated given the lag between the trigger event and subsequent observable AGN indicators).

\section{Conclusions}
In this work, we have investigated the Ca H plus H$\epsilon$ to Ca K ratio for cluster galaxies belonging to Abell~3888. Our work:

(1) demonstrates a new method to compute the H:K ratio that uses the continuum in between the Ca H and K lines as a reference point to determine the dips of these lines from. This in turn provides an indication of the relative strength and significance of the measured dips;

(2) shows the mean H:K value in Abell~3888, and by extension clusters more generally, is systematically lower than for mean values reported for the general galaxy population;

(3) shows that H:K ratios do not have a straight forward variance with either EW([O{\sc ii}]) or EW(H$\delta$). 

(4) shows that there is no preferred location for galaxies with inverted H:K ratios (H:K$>1.1$) within Abell~3888. Their dispersion values in the cluster phase-space are slightly lower than for strongly star-forming galaxies;

(5) suggests that strong H:K inversion galaxies are an heterogeneous population with a plurality of origin (e.g.\ merging star-forming spirals; quiescent galaxies merging with a gas-rich, low mass galaxy). 

Looking to the near future, we anticipate that the next public data release from SkyMapper (e.g., Wolf et al.\ 2018) may have good coverage of this region, as will, slightly later, the Large Synoptic Survey Telescope (cf.\ LSST Science Collaboration et al.\ 2009) . Both of these should yield superior imaging to Pimbblet et al.\ (2001) for these galaxies and help pin down further any physical properties that correlate with H:K inversion. 

\section*{Acknowledgements}
We wish to express our gratitude to the referee, Guy Worthey, for insightful feedback on the earlier draft of this work that has improved its content. We also thank our editor, Joop Schaye, for additional feedback on the manuscript.

We would like to express our gratitude to Scott Croom for convsersations and discussion on the finer points of the older 2dF instrumentation optics.

KAP acknowledges the support of STFC through the University of Hull Consolidated Grant ST/R000840/1.

%%%%%%%%%%%%%%%%%%%%%%%%%%%%%%%%%%%%%%%%%%%%%%%%%%



%\bibliographystyle{mnras}


%%%%%%%%%%%%%%%%%%%%%%%%%%%%%%%%%%%%%%%%%%%%%%%%%%

%%%%%%%%%%%%%%%%% APPENDICES %%%%%%%%%%%%%%%%%%%%%
%
%\appendix

% Don't change these lines
\bsp	% typesetting comment
\label{lastpage}
\end{document}